\documentclass[12pt]{article}
\usepackage{xspace}
\usepackage{epsfig}
\newcommand{\NPB}[3]{\emph{ Nucl.~Phys.} \textbf{B#1} (#2) #3}   
\newcommand{\PLB}[3]{\emph{ Phys.~Lett.} \textbf{B#1} (#2) #3}   
\newcommand{\PRD}[3]{\emph{ Phys.~Rev.} \textbf{D#1} (#2) #3}

\newcommand{\HPA}[3]{\emph{ Helv.~Phys.~Acta} \textbf{#1} (#2) #3}

\newcommand{\JETP}[3]{\emph{ JETP Lett.} \textbf{#1} (#2) #3}  

\newcommand{\nc}{\newcommand}

\def\simlt{\stackrel{<}{{}_\sim}}
\def\simgt{\stackrel{>}{{}_\sim}}

\nc{\beq}{\begin{equation}}
\nc{\eeq}{\end{equation}}
\nc{\beqa}{\begin{eqnarray}}
\nc{\eeqa}{\end{eqnarray}}
\nc{\bea}{\begin{eqnarray}}
\nc{\eea}{\end{eqnarray}}
\nc{\ra}{\rightarrow}
\nc{\lsim}{\begin{array}{c}\,\sim\vspace{-21pt}\\< \end{array}}
\nc{\gsim}{\begin{array}{c}\sim\vspace{-21pt}\\> \end{array}}

\nc{\LL}{L}
\nc{\vv}{\tilde{v}}

\title{
\vspace*{-1.3cm}
\begin{flushright}
\normalsize{
ANL-HEP-PR-04-27 \\
EFI-04-08 \\
FERMILAB-PUB-04/034-T
}
\end{flushright}
\vspace{1.5cm}
\Large
\textbf{ Dark Matter, Light Stops \\
and Electroweak Baryogenesis}
\vspace*{1.0cm}
\author{\large\textbf{C. Bal\'azs$^a$}, \textbf{M.~Carena$^b$}
and \textbf{C.E.M.~Wagner$^{b,c}$}\\ \\[0.5cm]
$^a$\normalsize\emph{HEP Division, Argonne National Laboratory,
9700 S. Cass Ave.,
Argonne, IL 60439, USA} \\
$^b$\normalsize\emph{Fermi National Accelerator Laboratory,
P.O. Box 500, Batavia, IL 60510, USA} \\
$^c$\normalsize\emph{Enrico Fermi Institute, Univ. of Chicago, 5640
Ellis Ave., Chicago, IL 60637, USA}}}

\begin{document}
\setcounter{page}{0}
\maketitle
%\date{}
\vspace*{1cm}
\begin{abstract}
We examine the neutralino relic density in the presence of a light top squark, 
such as the one required for the realization of the electroweak baryogenesis 
mechanism, within the minimal supersymmetric standard model. We show that there 
are three clearly distinguishable regions of parameter space, where the relic 
density is consistent with  WMAP and other cosmological data. These regions are 
characterized by annihilation cross sections mediated by either light Higgs 
bosons, Z bosons, or by the co-annihilation with the lightest stop. Tevatron 
collider experiments can test the presence of the light stop in most of the 
parameter space. In the co-annihilation region, however, the mass difference 
between the light stop and the lightest neutralino varies between 15 and 30 GeV, 
presenting an interesting challenge for stop searches at hadron colliders. We 
present the prospects for direct detection of dark matter, which provides a 
complementary way of testing this scenario. We also derive the required 
structure of the high energy soft supersymmetry breaking mass parameters where 
the neutralino is a dark matter candidate and the stop spectrum is consistent 
with electroweak baryogenesis and the present bounds on the lightest Higgs mass.
\end{abstract}

\thispagestyle{empty}
\newpage

\setcounter{page}{1}

\baselineskip18pt

%-------------------------------------------------------------
\section{Introduction}
%-------------------------------------------------------------

The questions of dark matter and baryogenesis lie at the interface 
between particle physics and cosmology. 
Recently, there has been an improved determination of the allowed range of cold
dark matter density from astrophysical and cosmological data. The Wilkinson 
Microwave Anisotropy Probe (WMAP) \cite{Spergel:2003cb}, in agreement with 
the 
Sloan Digital Sky Survey (SDSS) \cite{Tegmark:2003ud}, determined the matter 
and 
baryon density of the Universe to be $\Omega_m h^2=0.135^{+0.008}_{-0.009}$ and 
$\Omega_b h^2=0.0224\pm0.0009$, respectively, with $h=0.71^{+0.04}_{-0.03}$.
The difference between the matter and baryonic densities fixes the energy 
density of the cold dark matter as
\begin{equation}
\Omega_{CDM}h^2=0.1126^{+0.0161}_{-0.0181},
\end{equation}
at 95\% CL.  Here $\Omega_{CDM}$ is the ratio of the dark matter 
energy density to the critical density $\rho_c = 3 H_0^2/(8 \pi G_N)$, 
where $H_0 = h \times 100~km/s/Mpc$ is the present value 
of the Hubble constant, and $G_N$ is Newton's constant.
Such a  precise range of values 
poses important restrictions to any model of physics beyond the
Standard Model (SM) which intends to provide an explanation to 
the origin of dark matter.

Understanding what the observed dark matter in the Universe 
is made of is one of the most important
challenges of both particle and astroparticle physics, and collider
experiments are  an essential tool towards solving the dark matter puzzle.
Although there are many scenarios to explain the origin of dark matter,
weakly interacting massive
particles (WIMPs), with masses and interaction cross sections characterized
by the weak scale, provide the most compelling alternative. These neutral, 
and stable particles appear naturally
in low energy supersymmetry models, in the presence of $R$-parity 
\cite{reviews}. 
In a way, dark matter by itself provides
a fundamental  motivation for new physics at the electroweak scale. 
Low energy supersymmetry provides an excellent solution to the origin of dark 
matter and it has been extensively studied in the literature in different 
scenarios of supersymmetry breaking~\cite{Ellis:2003dn,Ellis:2003si,
Arnowitt:2001be,Feng:2004mt,Feng:2004zu,Djouadi:2001yk,Lahanas:2003bh}.

In contrast, the origin of the matter-antimatter asymmetry is
more uncertain. The three Sakharov requirements~\cite{sakharov} 
for a dynamical origin of the baryon asymmetry may be easily fulfilled in 
scenarios associated with the decay of heavy, weakly interacting
particles. Leptogenesis~\cite{leptogen} is an ingenious mechanism that 
explains the baryon asymmetry as induced
from a primordial lepton asymmetry which transforms
into a baryon asymmetry through weak anomalous processes. This is a very 
attractive scenario which yields a connection between baryon
asymmetry and neutrino physics. The heavy decaying 
particle may be identified with the lightest
right-handed component of the observable left-handed neutrinos.

One of the drawbacks of scenarios of baryogenesis
associated with heavy decaying particles is that they are difficult
to test experimentally. The minimal leptogenesis scenario is
consistent with the see-saw mechanism for the generation of the
small neutrino masses with all light neutrino masses below 0.1 
eV~\cite{BBP03},   
 but little more can be said without making
additional model-dependent assumptions. In particular, the 
CP-violating phase associated with neutrino oscillations is only indirectly
related to the phases associated with the generation of the 
primordial lepton asymmetry. 

Electroweak baryogenesis~\cite{EWBGreviews},
on the other hand, provides a scenario
that relies only on weak scale physics, and therefore potentially
testable at present and near-future experiments. 
Perhaps the most attractive feature of
this mechanism is that it relies on anomalous baryon number violation
processes which are present in the Standard Model~\cite{anomaly}. At
temperatures far above the electroweak phase transition critical
temperature, these
anomalous processes are unsuppressed and, in the absence of any $B-L$
asymmetry, they lead to the erasure of any baryon or lepton number
generated at high energy scales~\cite{ckn}. These baryon number
violation processes are, instead, exponentially suppressed in the
electroweak symmetry broken phase, at temperatures below the
electroweak phase transition~\cite{sphalerons,sphalT}. 
The mechanism of electroweak baryogenesis may become effective if 
the baryon number violation processes in the broken phase
are sufficiently suppressed
at the electroweak phase transition
temperature.
This, in turn, demands a
strongly first order electroweak phase transition, 
\begin{equation}
v(T_c)/T_c \simgt 1,
\end{equation}
where $v(T_c)$ denotes the Higgs vacuum expectation value at the
critical temperature $T_c$.

The strength of the first order phase transition may be determined by studying
the Higgs effective potential at high temperatures. 
The Higgs vacuum expectation
value at the critical temperature is inversely proportional to 
the Higgs quartic coupling, directly related to the Higgs mass
squared. For sufficiently light Higgs bosons, a
first order phase transition may take place, 
induced by loop-effects  
of light bosonic particles, with masses of order of the weak scale, and
strong couplings to the Higgs field. 
The only such particles in
the SM are the weak gauge bosons and their couplings are not
strong enough to induce a first-order phase transition for 
any value of the Higgs mass~\cite{SMpt}.

The condition of preservation of the baryon asymmetry may be
easily fulfilled by going beyond the SM framework. Within the
minimal supersymmetric Standard Model (MSSM), a first order phase
transition is still induced at the loop-level. The relevant
bosons are the supersymmetric partners of the top quarks (stops),
which couple strongly to the Higgs field, with a coupling equal
to the top-quark Yukawa coupling. In addition, a light stop
has six degrees of freedom, three of color and two of charge,
enhancing the effects on the Higgs potential. Detailed 
calculations show that for the mechanism of baryogenesis to
work, the lightest stop mass must be smaller than the top quark mass
and heavier than about 120~GeV.
Simultaneously, the Higgs boson involved in the electroweak
symmetry breaking mechanism must be lighter 
than 120~GeV~\cite{CQW}--\cite{LR}.

A light stop is an interesting possibility, independently
of the question of electroweak baryogenesis. These states
were and are being searched for at LEP and the Tevatron collider in various
decay modes. 
The Tevatron reach for a light stop depends
on the nature of the supersymmetry breaking scenario, which
determines the decay properties of the lightest stop, and
also on the specific values of the light chargino and
neutralino masses ~\cite{LEPSUSYt1,TeVt1,Debajyoti,Berger:1999zt,Chou:1999zb}. 
In this work, we focus on the case in which the lightest
neutralino is stable, and provides a good dark matter candidate.
In such case, the Tevatron can find a light stop provided
its mass is smaller than about 200 GeV~\cite{joekon}, a region that
overlaps maximally with the interesting one for electroweak
baryogenesis. 

Within
the MSSM, the lightest Higgs boson mass is bounded to be
below about 135 GeV~\cite{Heinemeyer:1998jw}. 
This bound depends crucially on the
top spectrum and also on the value of the CP-odd 
Higgs boson mass, $m_A$, and
the ratio of the Higgs vacuum expectation values, $\tan\beta$.
Lighter stops, or values of $\tan\beta$ of order one would
push this bound to values closer to the present experimental
bound, 
\begin{equation}
m_h \simgt 114.4 \; {\rm GeV},
\end{equation}
which is valid for a Higgs boson with SM-like couplings to the weak
gauge bosons~\cite{Barate:2003sz}. 
Consistency of the present Higgs boson bounds with a light
stop demands $\tan\beta$ to be large, $\tan\beta \simgt 5$,
and the heaviest stop to have masses of order 1~TeV or larger.
In particular, the requirement of a light Higgs boson, with
mass smaller than 120~GeV, 
necessary for the realization of electroweak baryogenesis, is
naturally fulfilled within the MSSM with one stop lighter 
than the top-quark.

\section{Light stop and dark matter constraints}
\label{LightStop}

The requirement that the lightest supersymmetric particle 
provides the observed dark matter of the Universe
demands that it should be lighter than the light stop.
Assuming that the lightest
supersymmetric particle is the superpartner of the neutral gauge
or Higgs bosons, namely a neutralino, 
imposes strong constraints on the values of the gaugino
and Higgsino mass  parameters. For simplicity, within this
work we shall assume that the gaugino mass parameters are
related by the standard unification relations, which
translate at low energies to $M_2 \simeq 2 \; M_1$, 
where $M_2$ and $M_1$ are the supersymmetry breaking masses
of the weak and hypercharge gauginos respectively. 

Interestingly enough, electroweak baryogenesis demands not
only a light stop and a light Higgs boson, but light
charginos or neutralinos as well~\cite{improved}. In particular, 
values of the Higgsino
mass parameter $|\mu|$  and $M_2$ of the same order,
and smaller than about 300~GeV are required. 
The presence of a light stop, as required for electroweak
baryogenesis, and a consistency with the observed
relic density of the Universe imposes then interesting
constraints on the parameter space.  

The introduction of non-vanishing phases is 
required to fully address the parameter space consistent
with electroweak baryogenesis. The required values
of the phases in the chargino sector vary in a wide
range, $1 \simgt \sin \phi_{\mu} \simgt 0.05$, where 
$\phi_{\mu}$ is the relative phase between the gaugino and
the Higgsino mass parameters~\cite{improved}. 
Larger values of this phase are preferred for larger values 
of the chargino masses and of the CP-odd Higgs mass~\cite{improved}.  
Due to the uncertainties involved in the determination of the
baryon asymmetries, however, one cannot exclude phases an order
of magnitude smaller than the ones quoted above.  

Recently, there have been several studies of the effects on
the neutralino relic density associated
with CP-violating phases in the soft supersymmetry breaking 
parameters~\cite{Gomez:2004ek,Nihei:2004bc,Argyrou:2004cs}. 
For large values of the CP-violating phases,
these effects are of special relevance in the 
neutralino annihilation  cross section via s-channel Higgs bosons,
as well as on direct dark matter detection,
due to the CP-violating effects on the couplings and the 
CP-composition of the Higgs boson mass 
eigenstates~\cite{CPHiggs}. In general, however, similar
effects to the ones coming from CP-violating phases may be
induced by changes in the value of the soft supersymmetry
breaking parameters in the CP-conserving case.

In this article we limit our study 
to the case of vanishing CP-violating phases. We expect
the CP-conserving case to represent well 
the constraints that exist for 
the lower end of  values of the phases consistent with
electroweak baryogenesis. Moreover, the CP-conserving
case addresses the general question about the constraints
coming from requiring an acceptable neutralino relic
density  in the presence of a light stop, like the one
accessible at the Tevatron collider.  As we shall show,
there are relevant implications for stop searches at hadron
colliders in general and at the Tevatron in particular,
in the region in which the stop co-annihilates with the
lightest neutralino. In this region, assuming universal gaugino
masses at the GUT scale, the lightest neutralino becomes mainly
bino and the resulting annihilation cross section becomes 
weakly dependent on the CP-violating phases.
A detailed study of the CP-violating case demands a
calculation that includes non-vanishing phases in a self-consistent way.
Work in this direction is in progress~\cite{progress}.

In order to define our analysis, we assume that all
squarks other than the lightest stop are heavy, with 
masses of order  1 TeV, and study the constraints that arise
in the $\mu$--$M_1$ plane from the requirement of
consistency with current experimental bounds and an acceptable
dark matter density.
Since the change of the sign of $\mu$
produces little variation in the selected regions of parameter
space, we shall show our results only for positive values of $\mu$.
We present results for $\tan\beta = 10$ and 50. 
Let us stress that the CP-violating sources for the generation of baryon number 
tend to be suppressed for large values $\tan\beta$, and that values of 
$\tan\beta$ not much larger than 10 are preferred from those considerations 
\cite{improved}. For $\tan\beta = 10$, we 
set the first and second generation slepton masses to 250 GeV, to accommodate 
the measured muon anomalous magnetic moment within one standard 
deviation\footnote{The latest calculations show a difference between 
the experimental~\cite{E821} and SM central values of the muon anomalous 
magnetic moment that varies in the range of about 5 and 25
$\times 10^{-10}$ with slight preference toward the higher 
end~\cite{deTroconiz:2004tr,hepph0402206,hepph0312264,hepph0312250,
hepph0312226,hepph0308213}. The SUSY 
contributions to $(g-2)_{\mu}$ were evaluated 
using the code described in Ref.~\cite{Baer:2001kn}.}. 
No significant variation of the neutralino
relic density is obtained by taking the sleptons to be as heavy as
the squarks. 

The mechanism of electroweak baryogenesis and the present bounds on the lightest 
CP-even Higgs mass demand one light-stop and one heavy-stop in the spectrum. 
Consistency with precision electroweak data is easily achieved by demanding that 
the light stop is mainly right-handed.  The required right- and left-handed stop 
supersymmetry breaking parameters are $m_{\tilde U_3} \simeq 0$ and $m_{\tilde 
Q_3} \simgt 1$~TeV, respectively \cite{CQW2}.
A non-negligible value of $A_t$ is necessary in order to avoid
the Higgs mass constraints. On the other hand, large values
of $A_t$ tend to suppress the strength of the first order phase
transition. Quite generally, acceptable values of the Higgs mass
and of the phase transition strength are obtained for
$0.3 \simlt |X_t|/m_{\tilde Q_3} \simlt 0.5$.

Our choices of the fixed weak scale soft 
supersymmetry breaking parameters can be 
summarized as follows
\begin{eqnarray}
&& m_{\tilde U_3}=0,~~~ m_{\tilde Q_3}=1.5{\rm~TeV},
 ~~~ X_t=\mu/\tan\beta-A_t=0.7{\rm~TeV},
 \nonumber \\
&& m_{\tilde L_3}\approx m_{\tilde E_3}\approx m_{\tilde D_3}\approx 1{\rm~TeV},
 ~~~m_{\tilde Q_{1,2}}\approx m_{\tilde U_{1,2}}\approx m_{\tilde D_{1,2}}
  \approx 1.2{\rm~TeV},                     
 \label{eq:MSSMSoftPars} \\
&& M_2=M_1g_1^2/g_2^2,~~~ M_3\approx 1{\rm~TeV}.
 \nonumber
\end{eqnarray}

Computations of masses and couplings at the one-loop level are performed
numerically using ISAJET 7.69 \cite{Paige:2003mg}, to which we  
input the weak
scale soft parameters, listed in Eq.(\ref{eq:MSSMSoftPars}) and in the figure 
captions.
To compute the neutralino relic density, we used ISAReD, 
the computer code which was presented in Ref. \cite{Baer:2002fv}.  
This code agrees well with the public code MicrOmegas \cite{Belanger:2001fz}. 
The neutralino-nucleon spin-independent scattering cross sections 
are computed by
the method described in Ref. \cite{Baer:2003jb}.  

\begin{figure}[t]
\centerline{\epsfysize=10.0cm\epsfbox{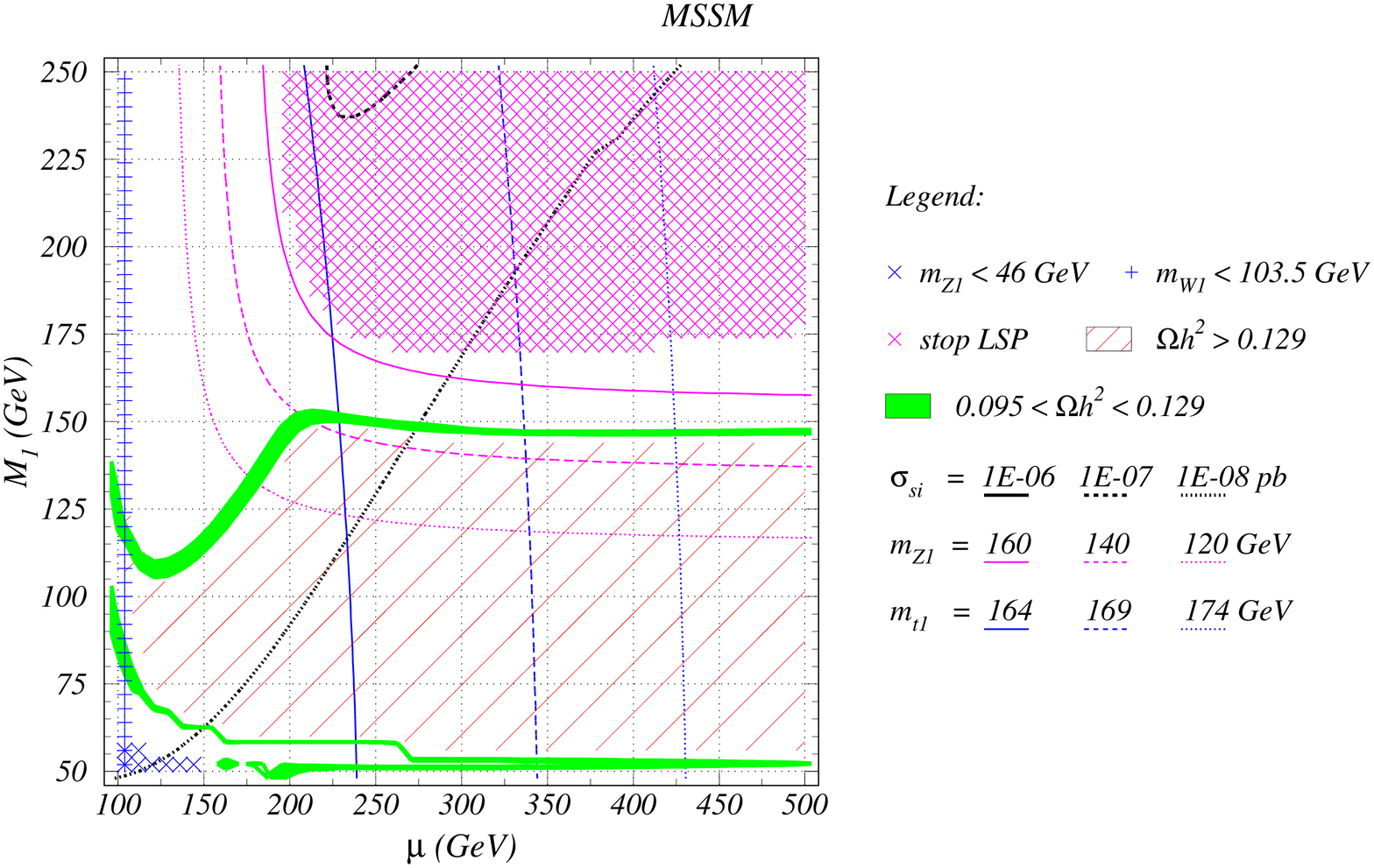}}
\caption{
Regions in the $\mu$--$M_1$ parameter space at which an acceptable value of cold 
dark matter develops.  The green bands show the region where the neutralino 
relic density is consistent with the WMAP data.  The black contours indicate 
cross section values for neutralino-proton scattering.  Neutralino and stop 
mass contours are also shown. Here we set $\tan\beta=10$, $m_{\tilde 
L_2}=m_{\tilde E_2}=250{\rm~GeV}$, and the CP-odd Higgs mass has been chosen to 
be equal to 500 GeV.}
\label{fig:mssm1}
\end{figure}
 
Figure \ref{fig:mssm1} shows the regions of parameter space, in the
$\mu$--$M_1$ plane, for which a consistent relic density develops in
the presence of a light stop and $\tan\beta = 10$. As we will show,
the allowed parameter space depends on the value of the 
CP-odd Higgs mass. In Figure~1, the 
CP-odd Higgs boson mass was
chosen to be 500 GeV and the
resulting  lightest CP-even Higgs mass lies
in the range 115--116~GeV.
The solid green (gray) area 
shows the region of parameter space where a neutralino relic density 
arises which is consistent with the WMAP observations at the 95\% CL.
The hatched regions are either incompatible with the neutralino
being the LSP (the neutralino becomes heavier than the stop) 
or excluded by LEP data \cite{LEPSUSYW1}. 
The regions of parameter space where the dark matter density is 
above the experimental upper bound and excluded by more than 
two standard deviations, 
are represented by the
shaded region in this figure. The white regions are
those where the neutralino relic density is below the 
experimental lower bound. An additional source of dark matter, unrelated
to the neutralino relic density, would be necessary in
those regions.  Constant stop and neutralino mass contours are
also shown by solid, dashed and dot-dashed curves, and are given
by approximate vertical and hyperbolic lines, respectively. 

\begin{figure}[t]
\centerline{\epsfysize=10.0cm\epsfbox{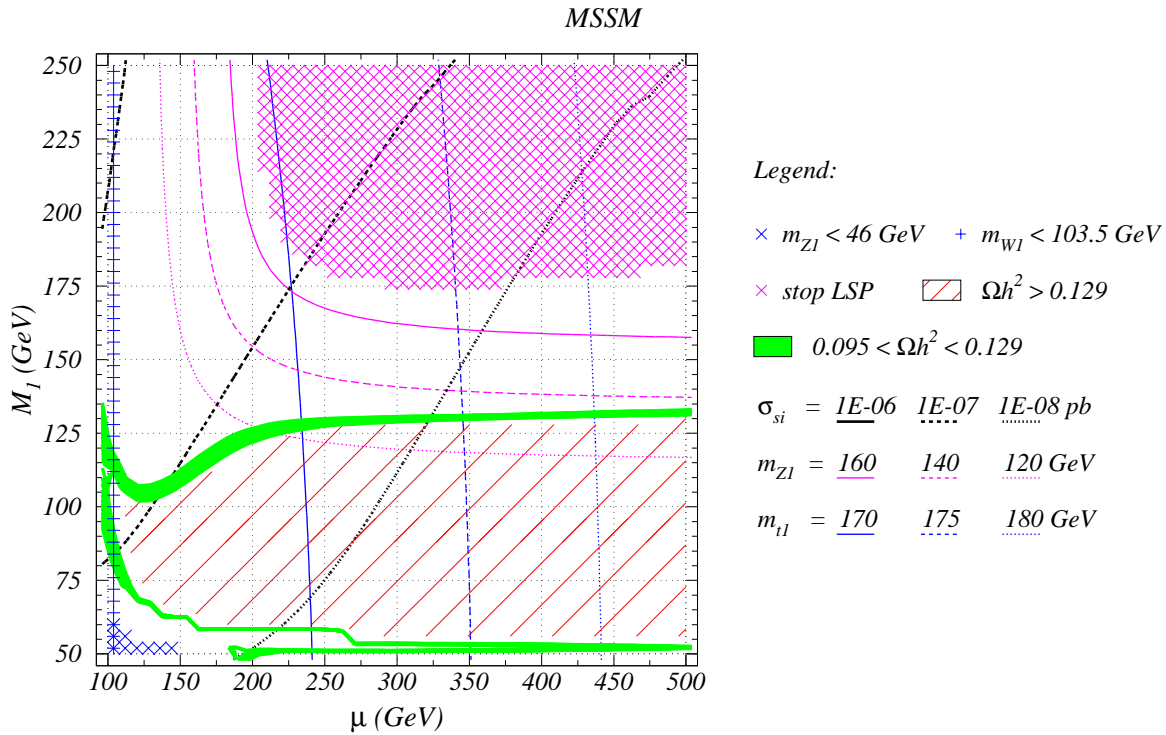}}
\caption{Same as Figure \ref{fig:mssm1}, except the CP-odd Higgs mass has 
been chosen to be equal to 300 GeV.}
\label{fig:mssm2}
\end{figure}

In Figure \ref{fig:mssm1}, 
there are three qualitatively
different regions in which an acceptable relic density arises. 
First, there is a region of parameter space where the mass
difference between the lightest neutralino and the light
stop is small. In this region, $M_1 \simeq 150$~GeV,
$|\mu| \simgt 200$~GeV,
stop-neutralino co-annihilation
dominates the neutralino annihilation cross section. The
stop-neutralino mass difference varies between 20 and 30 GeV
in that region, and, as we shall discuss below, presents 
a challenge for stop searches at hadron colliders. 

\begin{figure}[t]
\centerline{\epsfysize=10.0cm\epsfbox{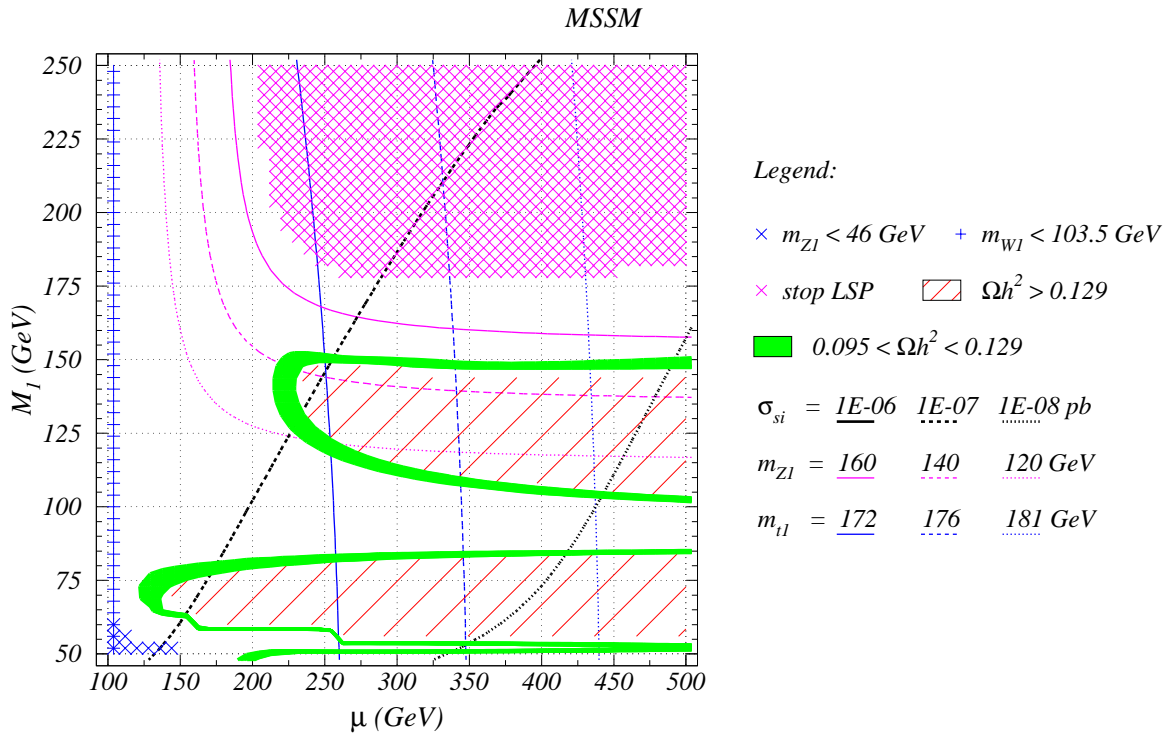}}
\caption{Same as Figure \ref{fig:mssm1}, except the CP-odd Higgs mass has 
been chosen to be equal to 200 GeV.}
\label{fig:mssm3}
\end{figure}

The second region of parameter space is the narrow band
present at small values of $M_1$ that becomes narrower
at large values of $\mu$. This region is associated with
the s-channel annihilation of the lightest neutralino
via the lightest CP-even Higgs boson. 
For large values of $|\mu|$ the value of 
$M_1$ at which this narrow band develops is approximately given by 
$M_1 \simeq m_h/2$.

The small width of the band at large values of $\mu$ may 
be explained by the fact that the Higgs-mediated
annihilation cross section is proportional to the square
of the small bottom Yukawa coupling. Indeed, for 
$m_A \simgt 200$ GeV and large values of $\tan\beta$,
the lightest Higgs boson has standard model like couplings
to all standard model fermions. The cross section is also
proportional to the square of the Higgs-neutralino coupling. 
This coupling is proportional
to both the gaugino and the Higgsino components
of the lightest neutralino. While for small values
of $\mu$ the neutralino may annihilate via Z-boson
mediated processes, no such annihilation contribution
exists for large values of $|\mu|$. Therefore, for
large values of $|\mu|$, 
the Higgs mediated s-channel  annihilation proceeds with a strength
proportional to the square of the Higgsino component, which 
decreases for large values of $|\mu|$, and inversely
proportional to the square of its mass difference
with the Higgs boson. That explains why the band
becomes narrower for larger values of $|\mu|$, for which 
a larger fine tuning between the Higgs and the 
neutralino masses is necessary in order to produce
the desired annihilation cross section.
Finally, there is a region for small values of
$|\mu|$ and $M_1$ for which the annihilation 
receives also contributions from Z-boson exchange
diagrams, which become rapidly dominant as the
neutralino mass is far from the stop mass or $m_h/2$.

In Figure \ref{fig:mssm1}, we also indicate cross section values for spin 
independent neutralino-proton scattering ($\sigma_{si}$).  Thick, 
black contour lines are plotted 
for  $\sigma_{si} = 10^{-7}$ pb (dashed) and 
$\sigma_{si} = 10^{-8}$ pb (dotted). 
For 
the parameters of Figure \ref{fig:mssm1}, these direct detection cross section 
values are close to $10^{-8}$ pb in the whole displayed region, with the 
exception of the low $\mu$ and high $M_1$ region, where $\sigma_{si} \sim 
10^{-7}$~pb. 
These cross sections are quite encouraging, since projections of GENIUS, 
XENON and ZEPLIN indicate future sensitivity even below $\sigma_{si} 
\sim 10^{-9}$ pb for the neutralino masses of interest \cite{CDMDetectors3}.  
As shown in the subsequent 
figures, for lower values of $m_A$ and/or higher values of $\tan\beta$,
direct detection cross sections occur up to $\sigma_{si} 
\sim 10^{-6}$~pb in the examined parameter region.  
These cross sections are at the reach of several experiments, 
such as CDMS, EDELWEISS and ZEPLIN.
According to their projections, these experiments will reach a sensitivity
of a few times $10^{-8}$~pb, in the next few years~\cite{CDMDetectors2}.

\begin{figure}[t]
\centerline{\epsfysize=10.0cm\epsfbox{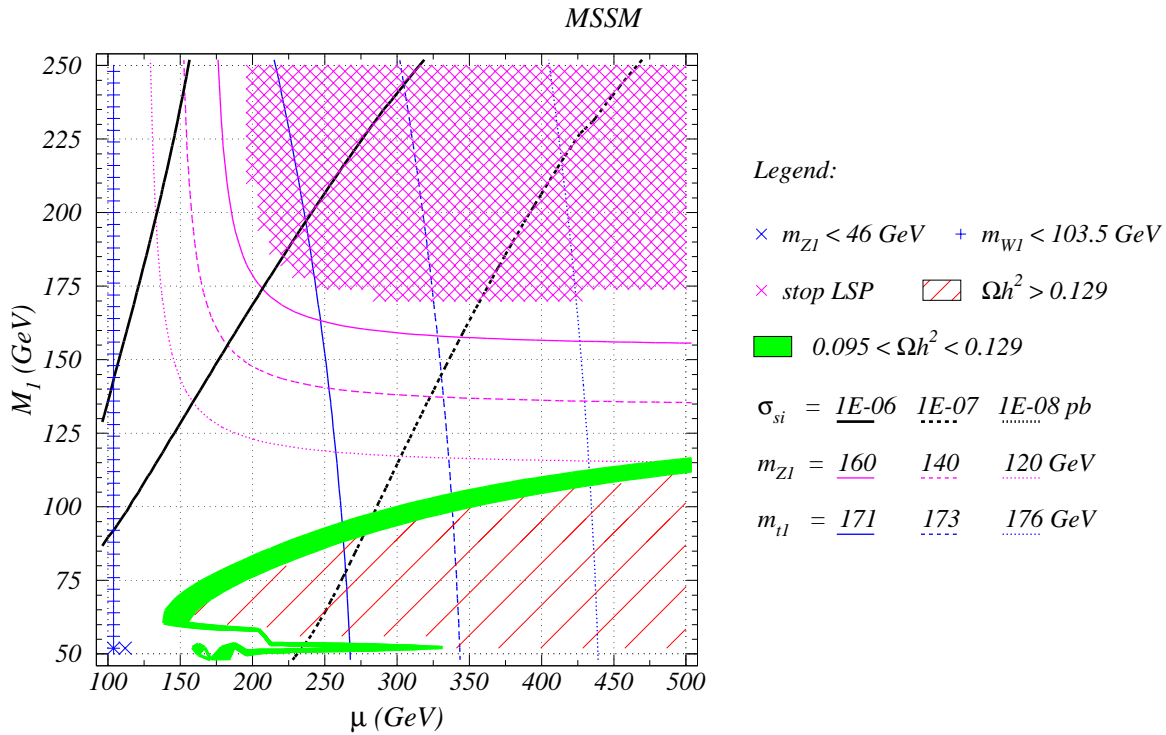}}
\caption{Same as Figure \ref{fig:mssm2}, except for $\tan\beta = 50$ and
$m_{\tilde L_2}=m_{\tilde E_2}=1.1{\rm~TeV}$.}
\label{fig:mssm501}
\end{figure}

So far, we have kept the CP-odd Higgs mass large, so
that the CP-odd Higgs has no impact on the annihilation
cross section. Figure 2 shows the case 
when the CP-odd Higgs mass is lowered to 300 GeV. For
this mass value, the CP-odd Higgs and the heavy CP-even
Higgs boson contribute
to the annihilation cross section in the s-channel 
resonant region for values of the neutralino mass
which are close to the stop mass. Due to the existence of
these two annihilation channels for similar values of the
neutralino mass, the main effect
of this smaller CP-odd Higgs mass is to move the region 
compatible with the observed
relic density slightly away from the stop-neutralino 
co-annihilation region to lower $M_1$ values.

\begin{figure}[t]
\centerline{\epsfysize=10.0cm\epsfbox{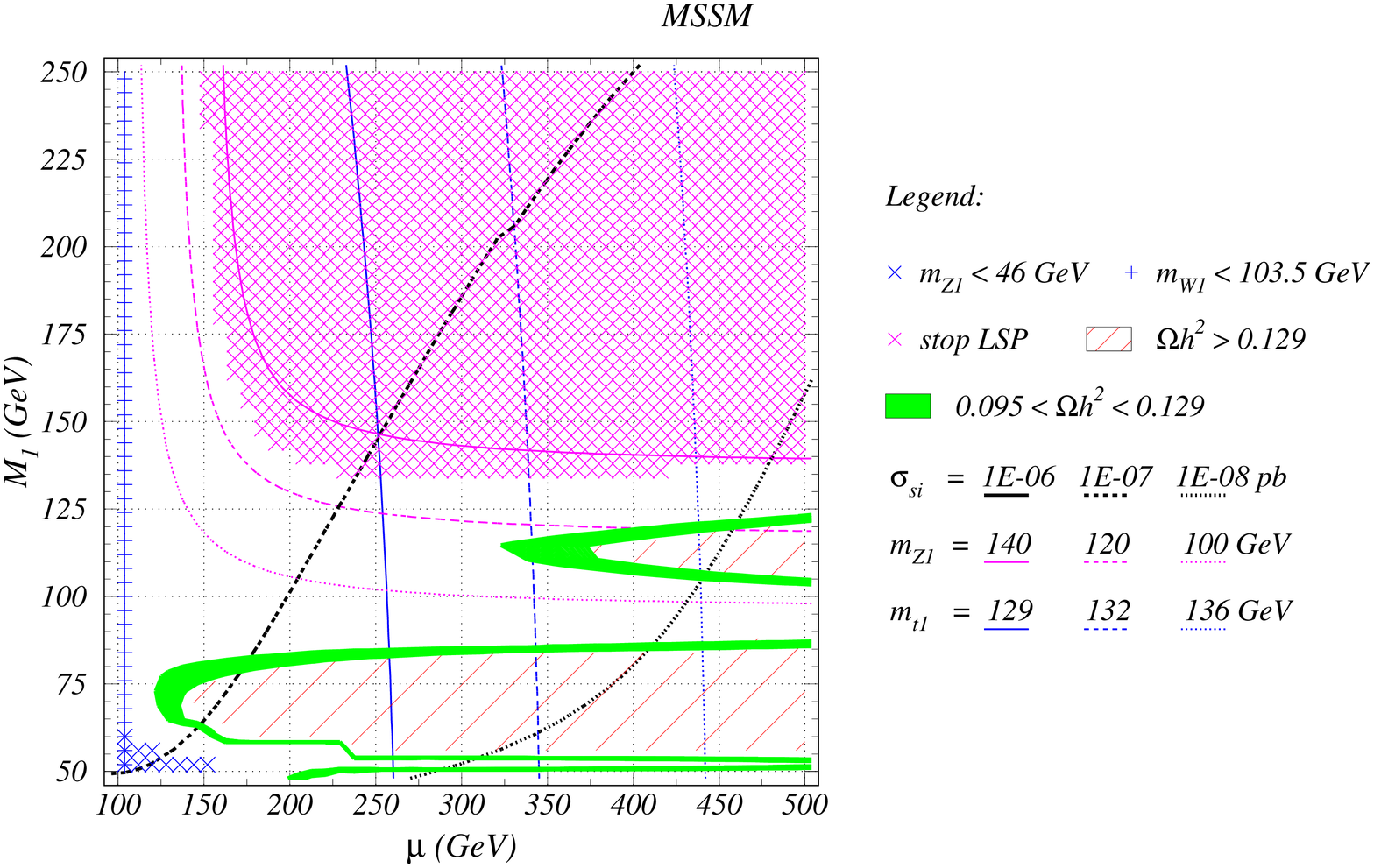}}
\caption{Same as Figure \ref{fig:mssm3}, except for 
$m_{\tilde U_3}=-(35$~GeV$)^2$ and $m_{\tilde Q_3}=1.25$~TeV.}
\label{fig:mssm29}
\end{figure}

In order to better visualize the importance of the
CP-odd Higgs mass affecting the stop-neutralino
mass difference in the region where co-annihilation is active,  
in Figure~\ref{fig:mssm3} we have plotted 
what happens when the CP-odd Higgs mass is moved towards
even smaller values, of order 200~GeV. In this case, the regions where
resonant annihilation via the CP-odd and heavy CP-even Higgs bosons
takes place are clearly shown.
There are now two bands of neutralino masses consistent with dark
matter density constraints, 
that appear at $M_1$
values of order 100 and 60 GeV, respectively, and that are associated
with s-channel annihilation via the lightest and heaviest Higgs
bosons, respectively. Observe that the width of the band associated
with s-channel annihilation via 
the CP-odd and heavy CP-even Higgs bosons becomes significantly 
larger than the ones appearing in Figure \ref{fig:mssm2}. 
This larger region is 
associated with the $\tan\beta$ enhanced couplings of these
particles to bottom quarks and tau leptons,
compared to the lightest CP-even Higgs boson.
In Figure~\ref{fig:mssm3},
the stop-neutralino co-annihilation region,
which had been modified in Figure \ref{fig:mssm2} due to the presence of the
300 GeV Higgs bosons, recovers the shape presented in Figure \ref{fig:mssm1},
with stop-neutralino mass differences of order 20-30 GeV.

Let us comment on the impact of varying the value of 
$\tan\beta$. The main effect is to change the coupling of the 
CP-odd Higgs boson and of the heavy CP-even Higgs boson to bottom
quarks. This coupling grows linearly with $\tan\beta$ and therefore
the annihilation cross section grows quadratically
with $\tan\beta$. The growth of the s-channel annihilation
cross section has dramatic consequences on the allowed parameter
space only if the CP-odd Higgs boson is light. Figure~\ref{fig:mssm501}
shows the impact of
taking $\tan\beta = 50$ for the CP-odd Higgs boson mass equal to 300~GeV.
While for $m_A = 500$~GeV, we obtain that
there is only a small variation with respect to the results of
Figure~\ref{fig:mssm1}, we show in Figure~\ref{fig:mssm501} that for 
$m_A = 300$~GeV the stop-neutralino mass gap becomes larger than the
one observed in Figure~\ref{fig:mssm2}. Finally, for $\tan\beta = 50$ and
$m_A = 200$~GeV, we obtain that the effect 
is sufficiently strong as to make the relic density smaller than
the observable one for most of the parameter space.
In Figure~\ref{fig:mssm501}, and 
for the computations with $\tan\beta = 50$, we have chosen 
heavy sleptons, to 
accommodate to the observed values of the muon anomalous magnetic
moment.
 
\begin{figure}[t]
\centerline{\epsfysize=10.0cm\epsfbox{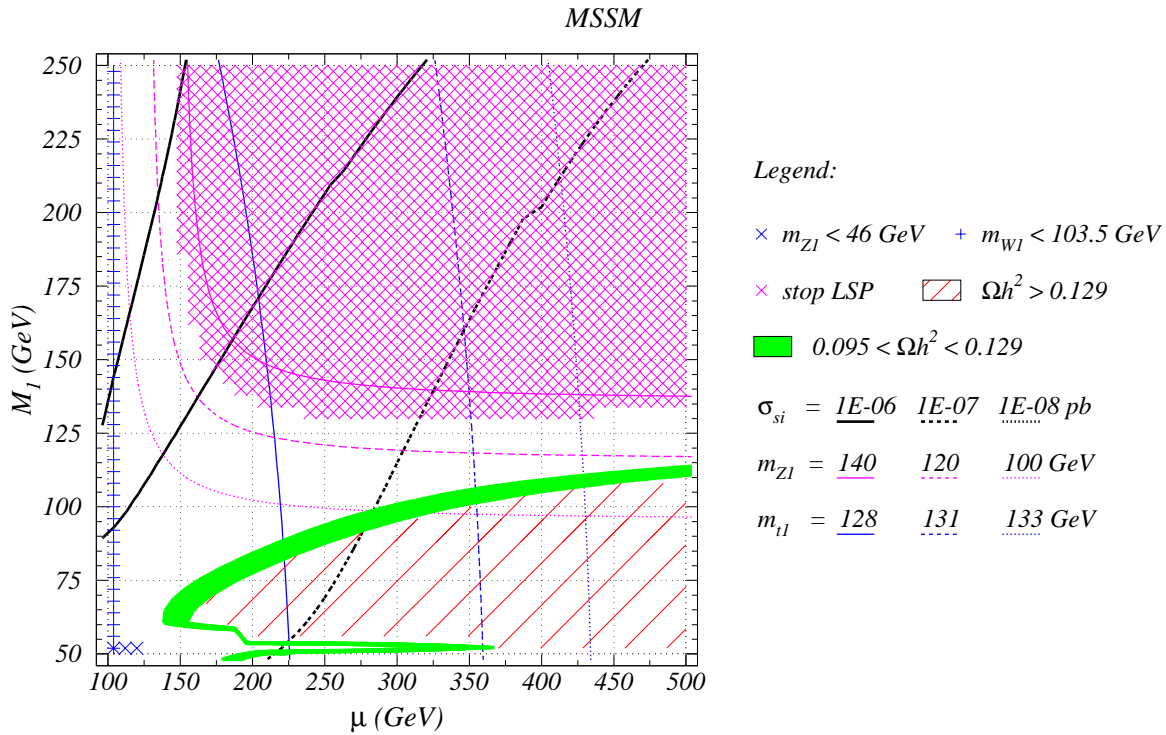}}
\caption{Same as Figure \ref{fig:mssm2}, except for $\tan\beta=50$, 
$m_{\tilde U_3}^2=-(35$~GeV$)^2$ and $m_{\tilde Q_3}=1.25$~TeV.}
\label{fig:mssm31}
\end{figure}

To further scrutinize the parameter region favorable for baryogenesis, we assigned 
a small negative value to the square of the right handed stop mass while 
simultaneously decreasing slightly the left handed doublet mass.  The results 
of the parameter scan are qualitatively the same as the ones presented in 
Figures \ref{fig:mssm1}-\ref{fig:mssm501}.  Figure \ref{fig:mssm29} shows a
representative case with $\tan\beta=10$ and $m_A = 200$~GeV.  Due to 
somewhat lower ${\tilde t_1}$ masses (in the region of 130 GeV) the 
co-annihilation region is squeezed to lower neutralino mass values.  The Higgs 
annihilation funnels are just the same as in Figure \ref{fig:mssm3}.  In 
contrast with the earlier results, the mass gap between the lightest stop 
and neutralino decreases to about 15 GeV.  

Finally, Figure \ref{fig:mssm31} shows a part of the parameter space with 
negative $m_{\tilde U_3}^2$ for $\tan\beta=50$.  For this value of $m_A$ (=300 
GeV) the co-annihilation and annihilation regions fuse just as in 
Figure~\ref{fig:mssm2}, but in the co-annihilation region the 
${\tilde t_1}$-${\tilde 
Z_1}$ mass gap remains about 15-20 GeV.  From this exercise of lowering the 
lightest stop mass, we draw the conclusion that for decreasing stop masses the 
stop-neutralino mass gap, which is necessary to satisfy the dark 
matter constraints with co-annihilation, also decreases. Although in this 
parameter region 
its mass is well within reach, with the smaller mass gap it is even more 
challenging for the Tevatron to discover the lightest stop.  On the other hand, 
the direct detection of relic neutralinos remains promising throughout this 
whole region.

In our analysis, we have not considered the constraints coming from the rare 
decay of $b\to s\gamma$.  In the absence of flavor violating couplings of the 
down squarks to gluinos this decay imposes a strong constraint on the Higgs and 
stop spectrum, in particular for large values of $\tan\beta$ \cite{bsg}. 
However, this constraint can be avoided in the presence of non–trivial down 
squark flavor mixing \cite{Gabbiani:1996hi,Carena:1998gk}.  For instance, 
assuming all down squark masses are of the order of 1 TeV, even a 
small left-right mixing of order of $10^{-2} \times m_{\tilde b}^2$
between the second and third generation down squarks can 
induce important corrections to the amplitude of this decay rate, that may 
compete with the one induced by the Higgs and the stop sectors.  This small 
mixing effects should have only a very small impact on our analysis of the high 
energy soft supersymmetry breaking parameters.

\section{Searches for a light stop at hadron colliders}

The search for a light stop in the MSSM depends both on the 
nature of the supersymmetry breaking mechanism as well as
on the mass difference between the light stop and the lightest
chargino and neutralino. When the mass difference between the
stop and the neutralino is small, the dominant decay channel
is a loop induced, flavor violating decay of the stop particle
into a charm and the lightest neutralino.
% \footnote{In models of 
% low energy supersymmetry breaking, such a decay
% will be subsequently followed by a decay of the neutralino
% into a gravitino and a photon (or a Z/Higgs boson), and 
% detection of such a decay process for stop masses up to
% 300~GeV is relatively easy. Tevatron collider with
% luminosity larger than 1 fb$^{-1}$ will essentially cover
% the whole parameter space consistent with such a 
% light stop~\cite{Debajyoti}.}

\begin{figure}[t]
\centerline{\epsfysize=10.0cm\epsfbox{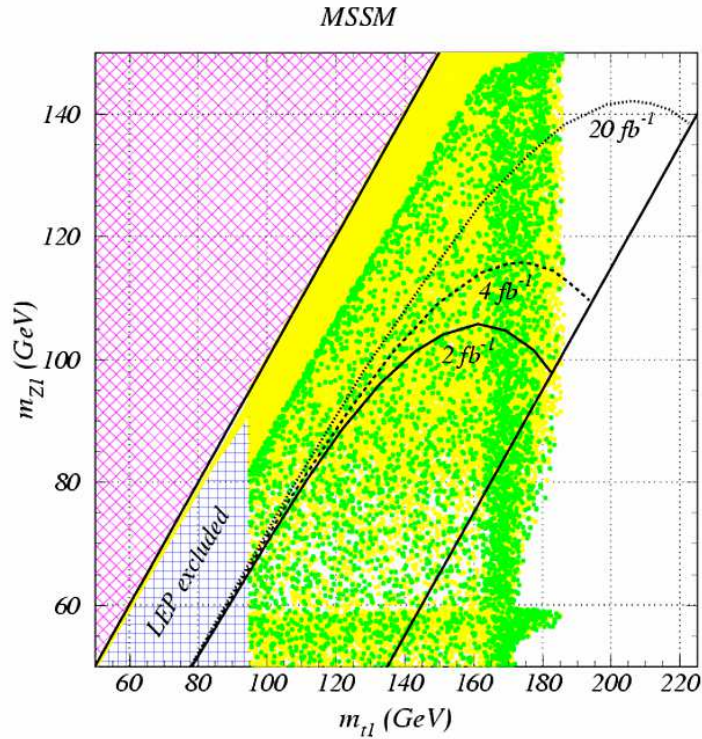}}
\caption{A random scan of parameter space projected on the stop mass vs. 
neutralino mass plane, as explained in the text in Eq.~\ref{eq:RandomScanPars}
and below.  The dark-gray (green) region is the one in which the relic density
is consistent with WMAP observations. In the light-gray (yellow) regions, the
relic density is below  the 2 $\sigma$ WMAP bounds. The hatched regions are
either excluded by LEP constraints (lower left) or inconsistent with the
assumption of a neutralino LSP.  
Overlayed the Tevatron light stop search sensitivity in the 
$cc\!\!\not\!\!{E_T}$ channel for 2 (solid), 4 (dashed) and 20 (dotted) pb$^{-1}$ 
integrated luminosity.}
\label{fig:mz1vsmst11}
\end{figure}

In models with a light stop and the neutralino providing the
observable dark matter, the neutralino signature will be
associated with missing energy ($\not\!\!\!{E_T}$). 
Detection of a decaying stop, that has a 
small mass difference with a lighter neutralino,
will depend on the ability of triggering on the missing
energy signature.
Present Tevatron search simulations for the region in which the two-body 
charm-neutralino decay is the dominant one are shown in 
Figure~\ref{fig:mz1vsmst11}~\cite{joekon}.
For 2--4 fb$^{-1}$ of
integrated luminosity and neutralino
masses smaller than 100~GeV, 
stops with masses up to about 180~GeV may be detectable under
the assumption that the stop-neutralino
mass difference is at least 30~GeV. Even
larger stop-neutralino mass differences are required for 
neutralino masses above 100~GeV, and stop detection becomes 
impossible for neutralino masses above 120~GeV.
(Some Tevatron limits are not shown in Figure~\ref{fig:mz1vsmst11},
since they are only effective for neutralino masses below 
50 GeV \cite{TeVt1}.)

In order to examine the stop-neutralino mass gap in the MSSM parameter space 
favorable for baryogenesis, we conducted a random scan over the following 
range of supersymmetric parameters:
\begin{eqnarray}
&& -(20 ~{\rm GeV})^2 < m_{\tilde U_3}^2 < 0, ~~~ 
    100 ~{\rm GeV} < \mu < 500 ~{\rm GeV}, ~~~
     50 ~{\rm GeV} < M_1 < 175 ~{\rm GeV}, \nonumber \\
&&  200 ~{\rm GeV} < m_A < 500 ~{\rm GeV}, ~~~
     10 < \tan\beta < 50.
\label{eq:RandomScanPars}
\end{eqnarray}
The rest of the parameters, which are not scanned, are fixed according to Eq. 
(\ref{eq:MSSMSoftPars}). 
The result of the scan, projected on the stop mass vs. neutralino mass plane, is 
shown by Figure~\ref{fig:mz1vsmst11}.  The (magenta $\times$) cross hatched area 
in the upper left corner is excluded since there the stop is lighter than the 
neutralino. Similarly, the (blue +) cross hatched region with stop masses below 
95 GeV shows the 95$\%$ C.L. exclusion limit of LEP \cite{LEPSUSYt1}.  
The green (dark gray) and 
yellow (light gray) dots represent models in which the relic density is 
consistent with or below the 2 $\sigma$ WMAP bounds.  (All these models also 
pass the $m_h > 114.4$ GeV and $m_{\tilde W_1} > 103.5$ GeV mass limits.) 

We concentrate on the green (dark gray) region of Figure~\ref{fig:mz1vsmst11}, 
where a relic density consistent with observations is obtained.
% For every neutralino mass,
% the region where co-annihilation with
% the lightest stop provides the dominant cross section for
% the determination of the relic density appears for the
% lightest stop masses. 
Neutralino co-annihilation with the lightest stop is dominant where the 
stop-neutralino mass gap is small.  
%This region is left to the (blue) 
%dash-dotted line which shows the $m_{\tilde t_1}/m_{\tilde Z_1} = 1.25$ mass 
%ratio.  
As it is apparent from the figure, under the present missing-energy 
triggering requirements, the Tevatron will not be able to detect a light stop 
in this region of parameters.

Away from the region where co-annihilation becomes efficient, the top searches 
depend strongly on the masses of the neutralinos and charginos.
As it is shown in the figure, prospects for stop detection improve dramatically 
so far the three body decay channel is suppressed by
\begin{equation}
m_{\tilde{t}} < m_{\chi^0} + m_W + m_b.
\label{condition}
\end{equation}

Searches at the Tevatron become more difficult for values of the 
stop and neutralino masses for which Eq. (\ref{condition}) is not
fulfilled. For stop masses above 140~GeV, this always happens in
the region of parameter space close to the one where the s-channel
annihilation  via the lightest CP-even Higgs becomes
efficient, which is clearly seen as a narrow band around
$m_{\tilde Z_1} \simeq$~58 GeV in Figure~\ref{fig:mz1vsmst11}.
In such a case, the three body decay mode 
becomes dominant, and the Tevatron, with less than
4 fb$^{-1}$ integrated luminosity, cannot detect
a light stop for any neutralino mass larger 
than 30~GeV~\cite{joekon}.

Searches for the stop may be complemented by searches
for charginos and neutralinos. An important channel
is the trilepton one, that will allow to test these
models for charginos masses up to about 130~GeV with
2 fb$^{-1}$. The value of the chargino mass in the 
gaugino region,
$|\mu| \simgt 200$~GeV, may be approximately identified
with the value of $M_2$, and using the standard
relation between $M_2$ and $M_1$, $M_1 \simeq M_2/2$, 
this implies values of $M_1$ smaller than about 
65~GeV~\cite{matchevpierce,sugra}. 
This covers the parameter 
space close to the region where s-channel annihilation
via the lightest CP-even Higgs boson becomes relevant,
in which stop searches become particularly difficult
when the stop is heavier than 140 GeV.

Finally, we comment on future searches at the LHC. While the LHC will
certainly be able to detect the charginos and neutralinos for all
of the parameter space consistent with dark matter and
a light stop, the search for a light stop may prove
difficult in the co-annihilation region for similar
reasons as at the Tevatron collider. Moreover,
the dominant stop production and detection channels at the
LHC come from the cascade decay of heavier colored
particles and it will be difficult to disentangle the
soft charm jets to identify the decaying top-squarks. 
A detailed study of LHC stop searches under these
conditions is required to draw firm conclusions.

\section{Direct dark matter detection}

Missing energy signatures at hadron- and lepton-colliders provide
very important evidence of the existence of a light, neutral, long-lived
particle in the spectrum. Within supersymmetric models, this particle
is identified with the LSP. But the stability of the lightest supersymmetric
particle on the time scales required to contribute
to the dark matter density cannot be checked by 
collider experiments. Direct detection of dark matter provides a 
complementary way of testing any particle physics explanation 
of the observed dark matter.

% The prospects for dark matter detection presented in 
% Figs.~\ref{fig:mssm1}--~\ref{fig:mssm31} rely on the
% standard assumptions of a dark matter flux incident on
% the earth, based on the observational evidence that points
% to a roughly spherical distribution of dark matter density
% in the galaxy, and a dark-matter velocity similar to the 
% speed of the sun within the galaxy.

Neutralinos of astrophysical origin are searched for in neutralino-nucleon 
scattering experiments detecting elastic recoil of nuclei.
The exclusion limits of these experiments are 
uniformly presented in the form of upper bounds on the (spin-independent) 
neutralino-proton scattering cross section ($\sigma_{si}$).  It is also 
customary to scale the cross section by a factor of 
\begin{eqnarray}
f = \left\{ \begin{array}{ll}
\Omega_{CDM} h^2/0.095 & {\rm if}~ 0.095 \geq \Omega_{CDM} h^2, \\
1                      & {\rm if}~ 0.095 < \Omega_{CDM} h^2 < 0.129,
\end{array} \right.
\end{eqnarray}
to account for the diminishing flux of neutralinos with their 
decreasing density \cite{Ellis:2000jd}.

\begin{figure}[t]
\centerline{\epsfysize=10.0cm\epsfbox{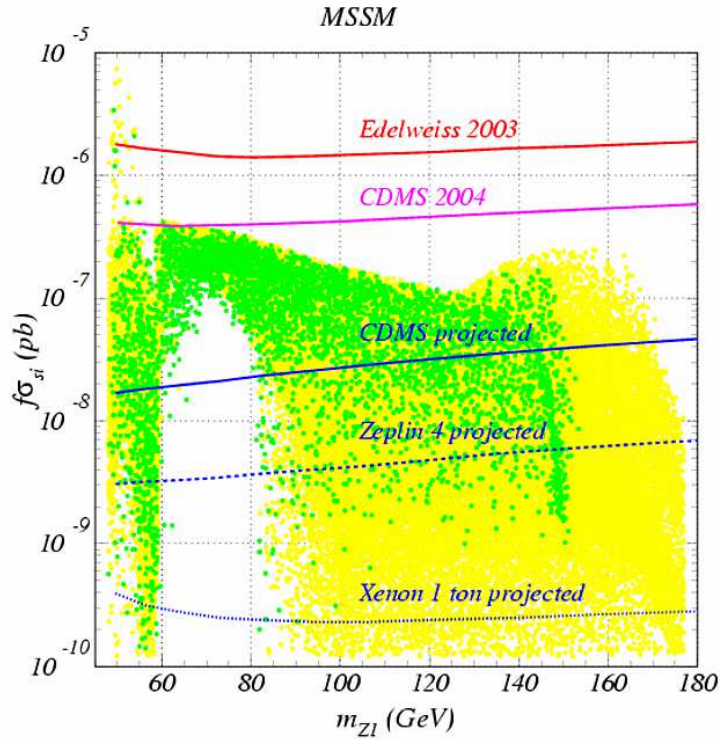}}
\caption{Spin independent neutralino-proton elastic scattering cross sections as 
a function of the neutralino mass. Green (dark gray) and yellow (light gray) 
dots represent models in which the neutralino density is consistent or below the 
2 $\sigma$ WMAP bounds. The top (red) and middle (magenta) solid lines represent
the 2003 and 2004 exclusion limits by EDELWEISS and by CDMS, respectively. 
The lower solid, dashed, and dotted (blue) lines indicate the projected 
sensitivity of CDMS, ZEPLIN and XENON, respectively.}
\label{fig:sigmasivsmz1}
\end{figure}

In Figure~\ref{fig:sigmasivsmz1}, we summarize the situation for dark matter 
detection in models with a stop lighter than the top quark, and the neutralino 
providing dark matter consistent with WMAP within 2 $\sigma$ (green dots). Here 
we use the result of the random scan over the range of SUSY parameters
defined by 
Eq.~(\ref{eq:RandomScanPars}).  For models marked by yellow (light gray) dots 
the neutralino relic density is below the 2 $\sigma$ WMAP bound, while models 
represented by green (dark gray) dots comply with WMAP within 2 $\sigma$.  
The top solid 
(red) line represents the 2003 exclusion limit by Edelweiss 
\cite{Chardin:2003vn},
while the middle solid 
(magenta) line shows the 2004 exclusion limit by CDMS 
\cite{Akerib:2004fq}.
The lower lines indicate the projected sensitivity of the 
CDMS (solid blue) \cite{Baudis:tv}, ZEPLIN (dashed blue) \cite{Cline:pi} and 
XENON \cite{Aprile} (dotted blue) experiments.
The `hole' that appears at neutralino masses around 60-80 GeV and cross 
sections below $10^{-7}$ pb is due to the LEP stop-mass excluded region.
This reflects the fact that small cross sections are induced only in the 
stop-neutralino co-annihilation region or in the resonant annihilation region
via the light CP-even Higgs for large value of $\mu$.

Prospects for direct detection of dark matter are quite good in most of the 
parameter space.  Presently the region above the (red) top and (magenta) middle 
solid line is excluded by EDELWEISS and by CDMS, respectively.  
But the CDMS (solid blue) and ZEPLIN (dashed blue) 
experiments will probe large part of the relevant parameter space.  
Finally, XENON (dotted blue) will cover most of the interesting region.

\section{High-energy soft supersymmetry breaking parameters}
\label{high}

The combined constraints of a light stop, as demanded for consistency
with the electroweak baryogenesis scenario, and an acceptable dark
matter relic density, imposes severe constraints on the soft
supersymmetry breaking mass
parameters of the theory. The stability of the lightest
super-partner implies that the gravitino is heavier than the lightest 
neutralino and in turn supersymmetry must be broken at 
high energies, and transmitted to the observable sector
at scales that are probably of order of the grand unification scale.
It is interesting to know which boundary conditions of the mass
parameters at high-energy scales could determine a low
energy spectrum consistent with a light stop and a light neutralino.

Neutralino dark matter in the presence of a light stop has been studied in 
Ref.~\cite{Boehm:1999bj,Ellis:2001nx} 
in the context of the minimal supergravity motivated 
model (mSUGRA).  Electroweak baryogenesis coupled with dark matter has also been 
examined in Ref.~\cite{davlos} within the same framework.  
It was found that for values
of $m_0$, and $M_{1/2}$ of order of the weak scale and much larger
values of $A_0$ ($|A_0| \simgt$ 1 TeV) it is 
possible to obtain acceptable neutralino relic density
in a narrow region of the  parameter space consistent with
electroweak baryogenesis.  Presently, with considerably 
stronger Higgs mass limits and WMAP constraints on the cold dark 
matter,  there is no mSUGRA parameter space where both dark matter 
and electroweak baryogenesis are satisfactory. 

Based on these considerations, 
we shall work under the assumption that the gaugino masses unify
at scales close to $M_{GUT} \simeq 10^{16}$~GeV, but we shall not 
assume unification of the scalar masses. Such boundary conditions
are natural in models of superstrings, where the values of the
supersymmetry breaking masses are determined by the vacuum expectation
value of the auxiliary components of dilaton and moduli 
fields~\cite{IbanezMunoz}.

For large values of $\tan\beta$, the mass parameter associated with
the Higgs acquiring the dominant vacuum expectation value is
of the order of the weak scale,
\begin{equation}
m_2^2 \simeq -\frac{M_Z^2}{2}.
\label{eq:m22}
\end{equation}
This mass parameter receives a supersymmetric contribution, proportional
to the square of the $\mu$ parameter, as well as one coming from the
supersymmetry breaking sector
\begin{equation}
m_2^2 = \mu^2 + m_{H_2}^2.
\label{eq:m22mu}
\end{equation}
Eqs.~({\ref{eq:m22}) and (\ref{eq:m22mu}) show that, in order to obtain
values of $|\mu|$ of the order of the weak scale, $m_{H_2}^2$ must be negative,
and of the order of the weak scale squared.

As discussed in Sec.~\ref{LightStop}, 
the realization of the mechanism of electroweak baryogenesis and the fulfillment
of the present bounds on the lightest CP-even Higgs mass 
require right- and left-handed stop supersymmetry
breaking parameters to be $m_{\tilde U_3} \simeq 0$ and 
$m_{\tilde Q_3} \simgt 1$~TeV, respectively.  
For values of $\tan\beta \simlt 20$, for which the 
bottom-quark Yukawa effects may be neglected,
the value of the low energy parameters are related
to the values at high energies by the approximate relation~\cite{COPW}
\begin{eqnarray}
m_{H_2}^2 & \simeq & m_{H_2}^2(0) + 0.5 M_{1/2}^2 +\Delta m^2,
\nonumber\\
m_{\tilde Q_3}^2 & \simeq & m_{\tilde Q_3}^2(0) + 7.1 M_{1/2}^2 + 
             \frac{\Delta m^2}{3},
\nonumber\\
m_{\tilde U_3}^2 & \simeq & m_{\tilde U_3}^2(0) + 6.7 M_{1/2}^2 + 
             \frac{2 \; \Delta m^2}{3}, 
\end{eqnarray}
where $M_{1/2}$ is the common gaugino mass at scales of order of
$M_{GUT}$, $m_i^2(0)$ denote the boundary condition of the scalar
mass parameters at this scale, and $\Delta m^2$ represents the 
negative radiative corrections governed by the large top-quark 
Yukawa coupling. Irrespectively of its form, the value of $\Delta m^2$ 
is related to $m_{\tilde U_3}^2(0)$ and $M_{1/2}^2$ 
by the relation
\begin{eqnarray}
\Delta m^2 & \simeq & - \frac{3 m_{\tilde U_3}^2(0)}{2} - 10 M_{1/2}^2
+ \frac{3 m_{\tilde U_3}^2}{2},
\nonumber\\
m_{\tilde Q_3}^2 & \simeq & m_{\tilde Q_3}^2(0) - \frac{m_{\tilde U_3}^2(0)}{2} + 3.1 M_{1/2}^2 
+ \frac{m_{\tilde U_3}^2}{2},
\nonumber\\
m_{H_2}^2 & \simeq & m_{H_2}^2(0) - \frac{3 \; m_{\tilde U_3}^2(0)}{2}
- 9.5 M_{1/2}^2 + \frac{3 m_{\tilde U_3}^2}{2}.
\end{eqnarray}

Additional information comes from the form of $\Delta m^2$. The
corrections depend on the square of the ratio of
the Yukawa coupling to its quasi-fixed point value~\cite{COPW}. 
For
values of $\tan\beta >5$, this ratio is close to two thirds and
one obtains, approximately
\begin{equation}
\Delta m^2 \simeq - \frac{ m_{\tilde Q_3}^2(0) + m_{\tilde U_3}^2(0) + m_{H_2}^2(0)}{3}
- \frac{A_0^2}{9} + 0.5 A_0 M_{1/2} - \frac{10}{3} M_{1/2}^2.
\end{equation}
The value of $M_{1/2}$ is related to the value of $M_1$ by
\begin{equation}
M_1 \simeq 0.4 M_{1/2}.
\end{equation}
Therefore, the value of $M_{1/2}$ varies from values of about 400~GeV
close to the stop-neutralino co-annihilation region, to values of
about 150~GeV, close to the region where s-channel annihilation
via the lightest CP-even Higgs boson becomes dominant.
For these values of $\tan\beta$, and $\mu$ of
the order of the weak scale, the low energy value of
$X_t \simeq -A_t$, with
\begin{equation}
A_t = \frac{A_t(0)}{3} - 1.8 M_{1/2}.
\end{equation} 
As discussed in Sec.~\ref{LightStop}, the 
acceptable values of the Higgs mass
and of the phase transition strength are obtained for
$0.3 \simlt |X_t|/m_{\tilde Q_3} \simlt 0.5$. 

Finally, the square of the CP-odd Higgs mass $m_A$ is approximately 
given by $m_1^2 + m_2^2$, or, approximately,
\begin{equation}
m_A^2 \simeq m_{H_1}^2(0) + 0.5 M_{1/2}^2 + \mu^2 - \frac{M_Z^2}{2}.
\label{eq:mH1ma}
\end{equation} 

Solving for the high energy parameters, we obtain,
\begin{eqnarray}
m_{\tilde U_3}^2(0) & \simeq & \frac{1}{3} \left(2 m_{\tilde Q_3}^2 - 2 \mu^2 
-M_Z^2 + 5 m_{\tilde U_3}^2 + 6 A_t^2 
+ 32 A_t M_{1} - 160 M_{1}^2 \right),
\nonumber\\
m_{H_2}^2(0) & \simeq & m_{\tilde Q_3}^2 - 2 \mu^2 
- M_Z^2 + m_{\tilde U_3}^2   + 3 A_t^2 
+ 15 A_t M_1 - 20 M_1^2,
\nonumber\\
m_{\tilde Q_3}^2(0) & \simeq & m_{\tilde Q_3}^2 - 0.5 m_{\tilde U_3}^2 
- 20 M_1^2 + 0.5  m_{\tilde U_3}^2(0),
\nonumber\\
A_t(0) & \simeq & 3 A_t + 13.5 M_{1}.
\label{highlow}
\end{eqnarray}

Eqs.~(\ref{eq:mH1ma}) and (\ref{highlow}) determine the high-energy value
of the soft supersymmetry breaking parameters in the Higgs and third
generation squark sector. For positive boundary conditions for the
mass parameters, small values of $m_A$ can only be
obtained for very small or vanishing values of $m_{H_1}^2(0)$, 
and small values of $|\mu|$. 
Positive values of $A_t$ tend to force all square mass
parameters $m_i^2(0)$ to be large, of the order of 1--2 TeV squared, 
and the value of $A_t(0)$ becomes extremely large, of about 3 to 4~TeV. 

For negative values of $A_t$, instead, the desired spectrum may be
obtained for values of $m_{\tilde U_3}^2(0) \simeq 0$ and values of $A_t(0)$ 
smaller than about 1~TeV. The values of the other two square-mass
parameters are of the same order, and of about a TeV-squared.
Therefore, a more natural high-energy--low-energy connection is
established for negative values of $A_t$. Using ISAJET
7.69~\cite{Paige:2003mg}, we checked numerically that these 
conclusions remain valid even after the inclusion
of higher loop RGE effects.

Finally, we mention that the main effect of increasing the
value of $\tan\beta$ is to add negative corrections, induced
by the bottom quark Yukawa coupling, to the parameters 
$m_{\tilde Q_3}^2$ and
$m_{H_1}^2$ at low energies. 
Therefore, for larger values of $\tan\beta$, the
required low energy spectrum would demand larger values
of $m_{\tilde Q_3}^2(0)$ and $m_{H_1}^2(0)$. 
In particular, light CP-odd Higgs bosons become more natural for large values
of $\tan\beta$.

\section{Conclusions}

The properties of the superpartners of the top quark have
an important impact on the determination of the Higgs mass
and also on the realization of the mechanism of electroweak
baryogenesis in the MSSM. Light stops may arise due to the
effect of mixing and also, as shown in section~\ref{high}, 
due to large negative
radiative corrections proportional to the square of the
large top-quark Yukawa coupling. 

In this article, we have studied the constraints that arise
on supersymmetric models once the presence of a light stop
and a consistent value of the dark matter relic density are
required. We have shown that there are three different
regions of parameter space in which these requirements may
be fulfilled, associated with different neutralino annihilation
channels. There are     regions of parameter space where
the s-channel annihilation into either Higgs bosons or 
Z-bosons become dominant, and appear for small values of
$M_1$ and large or moderate values of $|\mu|$, respectively.
The former regions depend strongly on the value of the 
CP-odd Higgs boson mass.

The presence of a light stop induces the existence of 
a third region, associated with co-annihilation between
the stop and the lightest neutralino. In such region of
parameters, unless the CP-odd Higgs boson is close to
twice the neutralino mass, the stop-neutralino mass 
difference tends to be smaller than 30 GeV, presenting
a serious challenge for stop searches at hadron colliders.
The prospects for stop detection at the Tevatron
collider away from this region of parameters remain promising.

While the Tevatron will explore an important region of
the MSSM parameter space compatible with electroweak
baryogenesis and the observed dark matter density, the
LHC will add in these searches by exploring the neutralino
and chargino spectrum and complementing the existing
Tevatron stop searches. 
The existence of the region where stop-neutralino
co-annihilation becomes dominant motivates a dedicated analysis 
of stop searches at the LHC, in the regions where the stop-neutralino
mass difference is small. 

Under the standard assumptions of neutralino density and velocity
distributions in our galaxy, prospects for direct dark-matter detection 
in the  coming years are very promising  in most of the MSSM parameter
space of interest in this work. 
Direct dark matter searches 
present a complementary way of testing  models of electroweak
baryogenesis and dark matter, beyond
the one provided by collider experiments.

Finally, we would like to emphasize that physics at a future linear 
collider~\cite{LC} would be very important to test
this scenario. On one hand, the LEP experience \cite{LEPSUSYt1} shows that 
lepton colliders have the potential to detect a stop even in the cases of small 
mass differences between the stop and the lightest neutralino. 
On the other hand, a linear collider will
provide the necessary precision to shed light on the nature and composition
of both the light stop and the dark matter candidate.

\section*{Acknowledgements}

Work at ANL is supported in part by the US DOE, Div.\ of HEP,
Contract W-31-109-ENG-38.
Fermilab is operated by Universities Research
Association Inc. under contract no. DE-AC02-76CH02000 with the DOE.
We thank G.~Servant for invaluable discussions, and M.~Schmitt and
T.~Tait for a critical reading of the manuscript.

\end{document}